\documentclass[%
 reprint, 
superscriptaddress,
 amsmath,amssymb,
 aps, prb,
]{revtex4-2}

\usepackage{graphicx}
\usepackage{dcolumn}
\usepackage{bm}
\usepackage{xcolor}
\usepackage{hyperref}
\hypersetup{
 pdfnewwindow=true, colorlinks=true,
 linkcolor=blue, anchorcolor=blue,
 citecolor=blue, filecolor=blue,
 menucolor=blue, urlcolor=blue}

\begin{document}

\preprint{APS/123-QED}

\title{Cubic BeB$_2$: A metastable $p$-type conductive material from first principles}

\author{Xiao Zhang}
\affiliation{Department of Materials Science and Engineering, University of Michigan, Ann Arbor, Michigan, 48109, USA}%
\author{Shashi Mishra}
\affiliation{Department of Physics, Applied Physics and Astronomy, Binghamton University-SUNY, Binghamton, NY 13902, USA}%
\author{Elena R. Margine }
\affiliation{Department of Physics, Applied Physics and Astronomy, Binghamton University-SUNY, Binghamton, NY 13902, USA}%
\author{Emmanouil Kioupakis}
 \email{kioup@umich.edu}
\affiliation{Department of Materials Science and Engineering, University of Michigan, Ann Arbor, Michigan, 48109, USA}%

\date{\today}

\begin{abstract}
Boron forms a wide variety of compounds with alkaline earth elements due to its unique bonding characteristics. Among these, binary compounds of Be and B display particularly rich structural diversity, attributed to the small atomic size of Be. Cubic BeB$_2$ is a particularly interesting phase, where Be donates electrons to stabilize a diamond-like boron network under high pressure. In this work, we employ \textit{ab initio} methods to conduct a detailed investigation of cubic BeB$_2$ and its functional properties. We show that this metastable phase is dynamically stable under ambient conditions, and its lattice match to existing substrate materials suggests possible epitaxial stabilization via thin-film growth routes.
Through a comprehensive characterization of its electronic, transport, and superconductivity properties, we demonstrate that cubic BeB$_2$ exhibits high hole concentrations and high hole mobility, making it a potential candidate for efficient $p$-type transport. In addition, cubic BeB$_2$ is found to exhibit low-temperature superconductivity at degenerate doping levels, similar to several other doped covalent semiconductors such as diamond, Si, and SiC.
\end{abstract}

\maketitle

\section{Introduction}

Boron is an element that holds a unique place within the periodic table. 
As a group III element, it forms III-V compounds with N, P, and As, whose functional properties have been consistently under investigation for decades. 
In addition, compounds formed between boron and electropositive elements from Groups I and II display diverse structures and electronic behaviors, owing to the unique bonding preferences of boron and the charge-donating role of the metals~\cite{VANDERGEEST2014stability}.
An example is MgB$_2$, one of the most extensively studied materials for high-temperature superconductivity~\cite{Nagamatsu2001,buzea2001review,kang2001mgb2}. 
In this compound,  electron donation from Mg enables the formation of a graphene-like B network. 
While MgB$_2$ has shown remarkable electronic properties, the relatively large size of Mg limits its ability to form more compact polymorphs with B, particularly those featuring 3D covalent boron networks. 
In addition, metal borides have been investigated as potential thermoelectric materials due to their ability to withstand high temperatures and relatively high $zT$ values~\cite{Saglik2023thermoelectric}. 

Among the boride materials, the compounds between Be, which is one row above Mg in the periodic table, and B, exhibit an interesting and complex behavior. 
The small atomic size of Be enables the formation of a wider variety of structural motifs in the Be-B compounds, allowing for more compact and tightly bonded configurations. 
Previous studies have identified a rich structural landscape for potential Be-B binaries, ranging from molecular to crystalline compounds, and spanning from B- to Be-rich compositions~\cite{hermann2013binary,hermann2012making,VANDERGEEST2014stability}.
From a chemical perspective, Be-B compounds with 1:2 stoichiometry are particularly interesting members among the Be-B binaries. 
The electronic configuration in BeB$_2$ enables the boron atom to behave similarly to a carbon atom, leading to structures that closely resemble those formed by carbon. 
This similarity has been demonstrated both at the crystalline level, where at least five metastable BeB$_2$ phases have been predicted theoretically~\cite{fan2014stable}, and in nanostructures, where carbon-like nanocrystals have been observed~\cite{Molina2005beb2}. 
Although the hexagonal phase of BeB$_2$, which is an MgB$_2$ analog, has not shown a superconductivity behavior as promising as MgB$_2$~\cite{profeta2001mgb2,nguyen2014new,FELNER2001absence}, the 2D form (monolayer) of BeB$_2$ has recently been studied as a potential anode material for battery applications~\cite{wan2022two,jia2017structural,YANG2024first}. 

\begin{figure}
    \centering
    \includegraphics[width=0.5\linewidth]{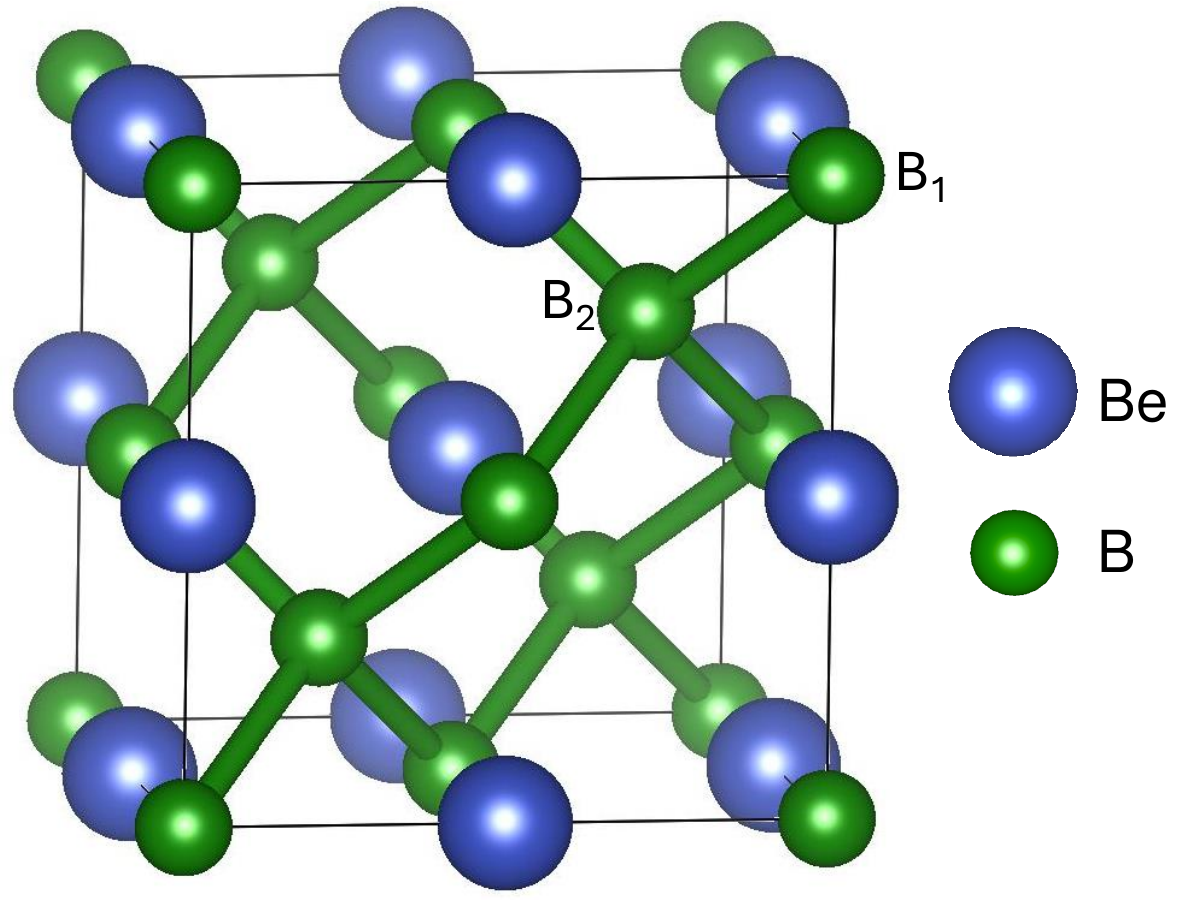}
    \caption{Atomic structure of $c$-BeB$_2$ (cF12, space group F4$\bar{3}$m, No.216), with Be donating two electrons to B, allowing B to form a network. The two B atoms are crystallographically inequivalent, with differing distances to the Be atom, giving the structure a zinc blende-like character rather than a homogeneous diamond-like network.
    The two boron atoms are marked by B$_1$ (closer to Be) and B$_2$ (farther from Be), respectively. 
    }
    \label{fig:structure}
\end{figure}

Among the various Be-B compounds, cubic BeB$_2$ ($c$-BeB$_2$) stands out due to its distinctive crystal structure. As shown in Fig.~\ref{fig:structure}, $c$-BeB$_2$ adopts a face-centered cubic (FCC) lattice with a Pearson symbol cF12 and belongs to the space group F4$\bar{3}$m (No. 216). 
The formation of this structure is made possible by the interplay between the small atomic size of Be and the valence states of both Be and B. In $c$-BeB$_2$, each Be atom donates two electrons to its neighboring B atoms, resulting in a four-electron valence configuration per B atom. 
The two B atoms in the unit cell are not equivalent due to the different distances from the Be atom, resulting in the formation of a zinc-blende-like network. Previous studies have identified $c$-BeB$_2$ as a metastable phase, thermodynamically stable only under high pressure conditions~\cite{fan2014stable,hermann2013binary,VANDERGEEST2014stability}, posing challenges for experimental synthesis. 
However, its unique bonding character warrants a comprehensive \textit{ab initio} investigation to explore its functional properties and the underlying structural-property relationships. 
The diamond-like crystal structure suggests high hole mobility, but in contrast to diamond, which has very high acceptor ionization energy, the metastable phase of BeB$_2$ is expected to have a high hole concentration, even intrinsic $p$-type behavior. 
Furthermore, inspired by the discovery of superconductivity in heavily doped diamond~\cite{ekimov2004superconductivity,Takano2004}, Si~\cite{bustarret2006superconductivity,Blase2009}, and  SiC~\cite{Ren2007,Kriener2008superconductivity}, the electronic structure of $c$-BeB$_2$, with boron atoms acting as pseudo-carbons, presents an opportunity for further exploration of its superconducting properties. 

In this work, we perform first-principles calculations based on density functional theory to investigate the electronic, optical, and superconducting properties of $c$-BeB$_2$. 
Our results confirm the dynamical stability of the cubic phase, and its close lattice match with existing substrate materials including 3C-SiC, MgO, and 4H-SiC, 
suggesting potential routes for thin-film stabilization. 
By evaluating the electronic structure, we show that the material exhibits a light effective hole mass, which favors $p$-type conductivity. 
We further examine the formation energies of several acceptor-type defects, and find that $c$-BeB$_2$ exhibits intrinsic $p$-type behavior with very low ionization energy. 
Simulations of electron transport predict a high intrinsic hole mobility exceeding 1,000~cm$^2$V$^{-1}$s$^{-1}$ at room temperature. 
The material is further found to be a low-temperature superconductor, with a critical temperature ($T_{\rm c}$) of approximately 3.6~K at a high doping level of $9.8 \times 10^{21} \, \mathrm{cm}^{-3}$. 

\section{Computational details}

The ground-state properties of $c$-BeB$_2$ are calculated with density functional theory using the Quantum Espresso (QE) package~\cite{giannozzi2009quantum,giannozzi2017advanced}. We employ optimized norm-conserving Vanderbilt (ONCV) pseudopotentials~\cite{schlipf2015optimization} with the Perdew-Burke-Ernzerhof (PBE) exchange-correlation functional~\cite{perdew1996generalized}, obtained from the PseudoDojo repository~\cite{vansetten2018the}. 
Structural optimization is performed through full relaxation of the unit cell, fitted to the Rose-Vinet equation of state~\cite{vinet1987temperature}. 
The resulting optimized structure has an external stress lower than $2\times10^{-5}$ Ry/Bohr$^3$. 
The wavefunctions are expanded as plane waves up to a cutoff energy of 100 Ry, and the Brillouin zone (BZ) is sampled with a $12\times12\times12$ grid, ensuring convergence of the total energy within 1 meV/atom. 

The electronic structure is obtained using the PBE exchange-correlation functional, the hybrid functional technique based on the Heyd-Scuseria-Ernzerhof (HSE06)~\cite{heyd2003hybrid,heyd2006erratum} scheme, as well as considering quasiparticle effects through many-body perturbation theory (the $GW$ approximation).
We evaluate the quasiparticle energies using the BerkeleyGW package~\cite{DESLIPPE2012berkeleygw,hybertsen1986electron}. 
A BZ sampling of $8\times8\times8$ is used to evaluate the quasiparticle energies. 
The dielectric screening is evaluated using the random-phase approximation, and its frequency dependence is approximated with the generalized plasmon-pole model of Hybertsen and Louie\cite{hybertsen1986electron}. The static remainder approach is used to treat the summation over empty bands\cite{DESLIPPE2012berkeleygw}. 
A cutoff of 40 Ry is used for evaluating the screened Coulomb interaction. 
Convergence of the quasiparticle calculations with respect to the number of summed empty states is shown in the supplemental information (Fig.S1)\cite{SM}. 
Subsequently, the maximally localized Wannier function (MLWF) technique~\cite{marzari2012maximally} is used to obtain a smooth electronic band structure. 
Hybrid functional calculations are performed with the VASP package~\cite{kresse1996efficient,KRESSE1996efficiency}, using the projector augmented wave (PAW) treatment of core electrons~\cite{kresse1999from}. 

To calculate direct optical properties, we solve the Bethe-Salpeter equation for optical polarization functions to take electron-hole Coulomb interactions into account~\cite{rohlfing2000electron}. 
A fine BZ sampling grid of $24\times24\times24$ is used to evaluate optical properties, with the energies and Coulomb matrix elements interpolated by considering wavefunction overlaps~\cite{DESLIPPE2012berkeleygw} from the coarse $8\times8\times8$ grid. 
The optical spectra of doped samples are evaluated by rigidly shifting the Fermi level due to the change in electron occupations, while neglecting effects such as partial ionization of dopants, or the modification of the band structure by the dilute dopant concentrations considered.

We use density functional perturbation theory (DFPT)~\cite{baroni2001phonons} to compute the dynamical matrix and the electron-phonon matrix elements on a $8\times8\times8$ BZ grid. The electron-phonon matrix elements are subsequently interpolated on a fine grid using the MLWF technique to enable accurate carrier transport calculations.
The hole mobility is determined by solving the Boltzmann transport equation, as implemented in the EPW code~\cite{Ponce2016epw,lee2023electron,ponce2021first}.
Both hole-phonon scattering and hole-ionized-impurity scattering are considered within the framework of the point charge model~\cite{leveillee2023ab} since in $p$-type semiconductors, the latter usually dominates at moderate to high doping levels. 
A fine grid of $120\times120\times120$ is used for transport calculations, and convergence tests confirm that increasing the grid density to $160\times160\times160$ results in mobility changes of less than 1\%. For superconductivity calculations, we use $80 \times 80 \times 80$ for $k$ and $q$-point grids, with an energy window of $\pm 0.3$~eV around the Fermi level. For calculating the isotropic Eliashberg spectral functions, the Dirac deltas of electrons and phonons are replaced by Gaussians of width 50~meV and 0.1~meV, respectively.
The zero-point and finite-temperature renormalization of the band gap due to electron-phonon interactions is evaluated using the Wannier function perturbation theory\cite{PhysRevX.11.041053} with $60\times60\times60$ phonon $\mathbf{q}$-grid, which converges the renormalization value within 0.3\%. 
The well-behaved convergence is consistent with recent results reported for diamond\cite{ponce2025verification}. 

In order to further evaluate defect formation in $c$-BeB$_2$, we calculate the total energy of pristine and defect-containing $2\times2\times2$ conventional supercells. 
The calculations are performed with the VASP package~\cite{kresse1996efficient,KRESSE1996efficiency}, using the PAW treatment of core electrons~\cite{kresse1999from} with the PBE exchange-correlation functional. 
To ensure accurate formation energies, we further relax the cells using the hybrid functional technique in the framework of HSE until the internal forces between atoms in the supercell are below 10 meV/\r{A}. 
Several acceptor type defects are considered to examine the $p$-type dopability of the material. The defect formation energy for a given charge as a function of the Fermi level is given by: 
\begin{equation}
    E_{f}^{\mathrm{defect}}(q) = E_{\mathrm{tot}}^{\mathrm{defect}} - E_{\mathrm{tot}}^{\mathrm{bulk}} - \sum_i n_i \mu_i + q(E_{\mathrm{F}} + E_{\mathrm{v}}) + E_{\mathrm{corr}}.
\end{equation}
$E_{\mathrm{tot}}^{\mathrm{defect}}$ and $E_{\mathrm{tot}}^{\mathrm{bulk}}$ are the total energies of the defect and pristine supercells, respectively. $n_i$ and $\mu_i$ denote the number of defect atoms and the corresponding chemical potential. $E_F$ and $E_v$ are the Fermi level and the energy of the valence band maximum (VBM), respectively. 
$q$ represents the charge state of the defect. 
$E_{corr}$ is a correction induced by considering periodic charge, and we adopt the scheme of Freysoldt et al.~\cite{freysoldt2009fully}, with the additional corrections due potential alignment evaluated using the \textit{pymatgen-analysis-defects} package~\cite{shen2024pymatgen}. 
The ranges of elemental chemical potentials are determined  by the formation enthalpies of $c$-BeB$_2$ and the two secondary phases associated with the dopants considered, LiB and BH$_3$, evaluated using the HSE06 exchange-correlation functional.

\section{Electronic, vibrational, and optical properties}

\begin{figure}[!b]
    \centering
    \includegraphics[height=4.5 cm]{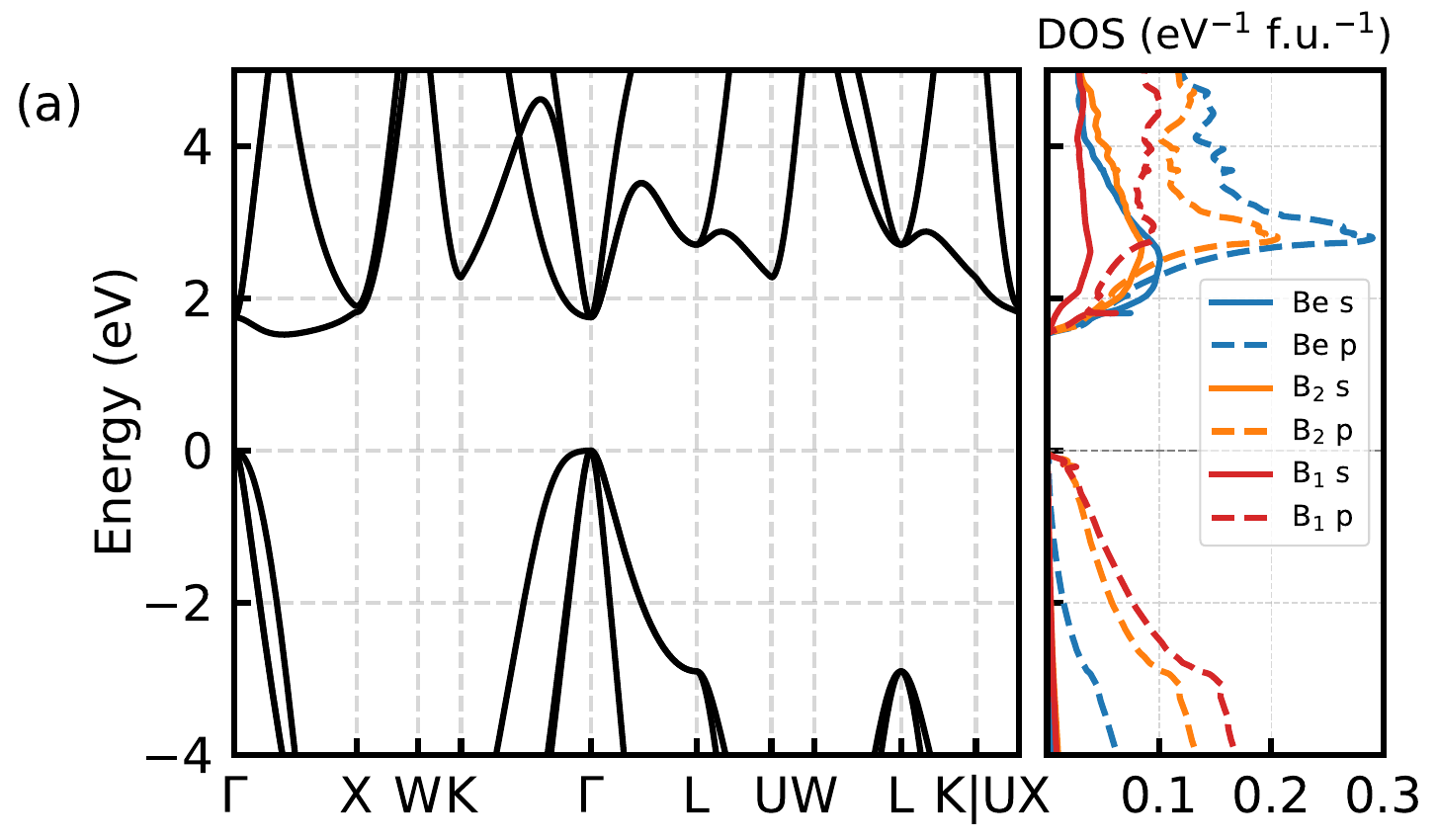}
    \includegraphics[height=4.5 cm]{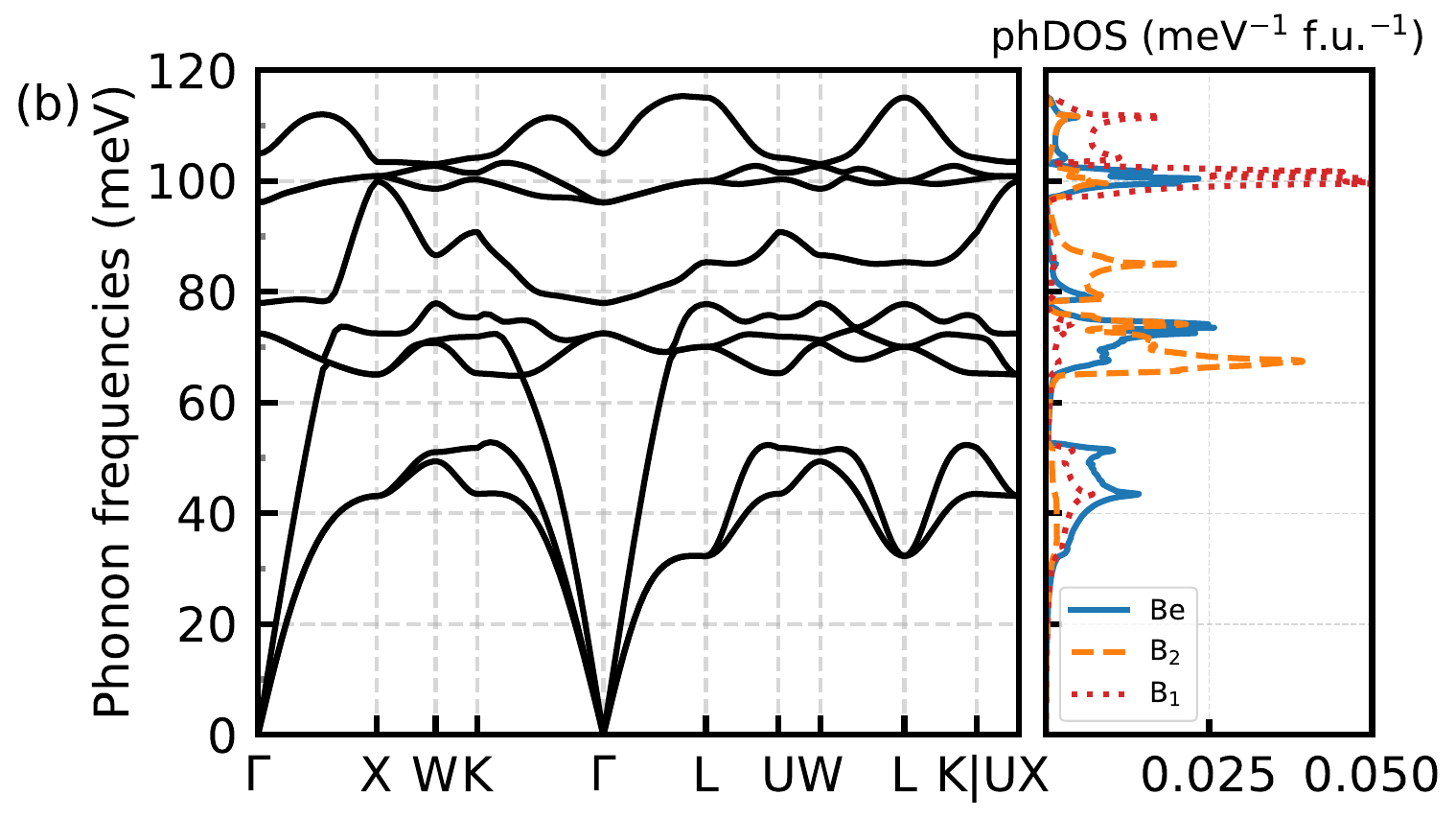}
    \caption{(a) Electronic band structure of $c$-BeB$_2$ evaluated within the HSE06 exchange-correlation functional. With HSE06, the material is found to be an indirect gap semiconductor with a fundamental gap of 1.52 eV. (b) Phonon dispersion indicating dynamical stability of $c$-BeB$_2$. }
    \label{fig:bands}
\end{figure}

The relaxed lattice parameter of $c$-BeB$_2$ is found to be $a=4.315$ \r{A} with the PBE exchange-correlation functional, and $a=4.297$ \r{A} with the HSE hybrid functional. 
The calculated B$-$B bond length is 1.86 \r{A}, slightly longer than the C$-$C bond length in diamond (1.6 \r{A} from simulation and 1.55 \r{A} from experiment~\cite{calzaferri1996band}).
The bond length of Be$-$B$_1$ is 1.86 \r{A}, and the bond length of Be$-$B$_2$ is 2.15 \r{A}. 
The formation enthalpy is evaluated as $\Delta H_f=E_{\text{BeB$_2$}}-E_{\text{Be}_{hcp}}-E_{\text{B}_\alpha}$, yielding a value of $-$0.018 eV/atom ($-$0.054 eV/f.u.) with the HSE06 hybrid functional, in excellent agreement with the literature value of $-$0.017 eV/atom~\cite{fan2014stable,hermann2013binary}. 
An early study on Be-B compounds reported $c$-BeB$_2$ as the ground-state structure at this composition under ambient pressure~\cite{hermann2013binary}. However, it only lies on the convex hull above 80 GPa~\cite{hermann2013binary}, due to the presence of the thermodynamically stable BeB$_{2.75}$ phase at ambient conditions. 
Subsequent work identified a lower-energy BeB$_2$ phase at ambient pressure that crystallizes in the orthorhombic structure (oS12, $Cmcm$\, No. 63)~\cite{VANDERGEEST2014stability,nguyen2014new,fan2014stable}, though this phase remains metastable with respect to the convex hull by 13 meV/atom. 
Consequently, $c$-BeB$_2$ (cF12) is a higher energy metastable phase, lying 85~meV/atom above BeB$_2$ (oS12) phase and 98~meV/atom above the convex hull~\cite{VANDERGEEST2014stability} at ambient pressure. 
We verified that the material remains dynamically stable and no significant changes in the electronic structure are seen at a very high pressure of 80 GPa (see supplemental materials Fig.~S2\cite{SM}). 
However, the significant energy offset suggests that bulk synthesis of $c$-BeB$_2$ is difficult to achieve experimentally, consistent with the experimental observation of a large uncertainty of phases in Be-B compounds~\cite{hermann2012making}. 
Despite the large energy barrier, the 4.297 \r{A} lattice constant of $c$-BeB$_2$ 
closely matches those of several common substrates, including 3C-SiC (4.359 \r{A}, -1.4\%)~\cite{feng2013sic} and MgO (4.212~\r{A}, 1.9\%)~\cite{lide2004crc}. 
Additionally, the equivalent lattice constant in the (111) hexagonal plane of $c$-BeB$_2$ is 3.038~\r{A}, which aligns well with the in-plane lattice constant of 4H-SiC (3.073~\r{A}, -1.1\%)~\cite{feng2013sic}.
As a result of this favorable lattice match to cubic systems, epitaxial stabilization can serve as a possible route for realizing this metastable phase in thin-film form. 
Importantly,  metastable cubic BeB$_2$  and the thermodynamically stable BeB$_{2.75}$ polytype (which lies on the convex hull)  adopt markedly different crystal structures. 
BeB$_{2.75}$ crystallizes in a large hexagonal unit cell (consisting of 110 atoms) that is composed of a complex network of icosahedral boron clusters ($P6/mmm$, space group No. 191). This structure is incompatible with high-symmetry cubic substrates, such as MgO, 3C-SiC. Moreover, its significantly larger hexagonal lattice constant ($a=9.774$ \r{A}\cite{CHAN2002385}) is mismatched by 5.7\% compared to $3\times$ the hexagonal lattice constant of 4H-SiC. Therefore, BeB$_{2.75}$ is less likely to nucleate coherently under epitaxial conditions on these substrates, with which cubic BeB$_2$ is nearly lattice matched.


\begin{table}[]
\small
    \centering
    \begin{tabular}{c|ccccc}
    \hline
         & PBE &HSE&$GW$ & Ref.\cite{fan2014stable}, PBE &Ref.\cite{hermann2013binary}, PBE \\ \hline
        E$_{\text{g,ind}}$ (eV) & 0.99 & 1.52 & 1.58 & 0.95 & 1.00 \\
        E$_{\text{g,dir}}$ (eV) & 1.27&1.75 & 1.86 &- &-\\
        \hline
    \end{tabular}
    \caption{Calculated electronic band gaps of $c$-BeB$_2$ using different methods. Our PBE result is in good agreement with literature values. 
    Both PBE+$GW$ and HSE06 yield very similar indirect band gaps. 
    A consistent difference of approximately 0.3~eV between the direct gap and indirect gap is found across all methods. 
    }
    \label{tab:gaps}
\end{table}

\begin{table*}[!ht]
\centering
\small
\caption{Hole effective mass in $c$-BeB$_2$ along different directions. The light holes have much lighter effective masses compared to those in typical FCC semiconductors.}
\label{tab:mass}
\begin{tabular}{c|cccccc}
\hline
Direction & $c$-BeB$_2$(PBE) & $c$-BeB$_2$(HSE) & Diamond\cite{lofas2011effective} & Silicon\cite{ponce2018towards} & cubic BAs\cite{bushick2019band} & cubic SiC\cite{willatzen1995relativistic} \\
\hline
$m_{lh, \Gamma\text{-}X}$ & 0.088 & 0.086 & 0.26 & 0.202 & 0.243 &0.470\\
$m_{hh, \Gamma\text{-}X}$ & 0.652 & 0.545 & 0.36 & 0.243 & 0.253 &0.662\\
$m_{lh, \Gamma\text{-}K}$ & 0.085 & 0.082 & 0.23 & 0.140 & 0.195 &0.339\\
$m_{hh, \Gamma\text{-}K}$ & 10.906 & 5.913 & 1.39 & 0.512 & 0.789 &1.46\\
$m_{lh, \Gamma\text{-}L}$ & 0.061 & 0.059 & 0.66 & 0.132 & 0.136 &0.321\\
$m_{hh, \Gamma\text{-}L}$ & 0.226 & 0.224 & 0.65 & 0.643 & 0.727 &1.91\\
\hline
\end{tabular}
\end{table*}

We further computed the electronic band structure and phonon dispersion of $c$-BeB$_2$, shown in Fig.~\ref{fig:bands}. The electronic band gap values at different levels of theory are listed in Table~\ref{tab:gaps}, along with comparisons to literature reports. 
A comparison of the band structure from PBE and HSE exchange-correlation functionals is shown in Fig.~S3~\cite{SM}.  
The band gap evaluated with the PBE exchange-correlation functional agrees well with previously reported values~\cite{fan2014stable,hermann2013binary}. 
Notably, the $c$-BeB$_2$ phase remains the only semiconducting phase among all known polytypes. 
An indirect band gap is found consistently using the PBE and HSE exchange-correlation functionals, as well as with single-shot $GW$ based on the PBE calculations. 
While PBE is known to severely underestimate the gap, the band gap from the HSE exchange-correlation functional (1.52 eV) and the $GW$ (1.58 eV) approximation are found to be very close. 
The difference between the indirect band gap, located along the $\Gamma-X$ direction, compared to the minimum direct gap, located at the $\Gamma$ point, is found to be approximately 0.23-0.28 eV.
The electron density of states evaluated from the HSE exchange-correlation functional shows that the valence band maximum is composed primarily of boron $p$ orbitals. Both boron atoms contribute comparably up to $\sim$ 0.5~eV below the VBM, below which the contribution from the B$_1$ atoms (which are closer to Be) becomes more dominant. 
The conduction band minimum, on the other hand, is mainly composed of $p$ states from the Be atoms and the B$_2$ atoms (the ones farther from Be). 
By analyzing the Born effective charges below, we provide a more detailed insight on the relationship of the bonding and the distance of the atoms. 

We evaluate the hole effective masses by performing parabolic fits to the electronic band structures calculated with both the PBE and the HSE exchange-correlation functionals. We consider states within 5\% of the VBM along the three high-symmetry directions of the BZ: $\Gamma-X$(100), $\Gamma-K$ (110), and $\Gamma-L$ (111). The resulting effective masses are listed and compared to literature values for other common cubic semiconductors in Table~\ref{tab:mass}. 
We find a pronounced difference between the light and heavy hole effective masses in $c$-BeB$_2$, more so than in materials such as SiC, diamond, and $c$-BAs. 
In particular, the light holes exhibit significantly lower effective masses across all directions, suggesting the potential for efficient hole transport in this material.  
The trend is consistent between PBE and HSE. 
Quantitatively, the inclusion of the Hartree exact exchange in HSE widens the bands near the VBM, therefore increasing the effective masses in general. 
However, the effective masses of the light holes remain similar along all directions, with the most pronounced impact of HSE on the heavy holes along the $\Gamma-K$ direction. 

\begin{figure}[!b]
    \centering
    \includegraphics[width=1.0\linewidth]{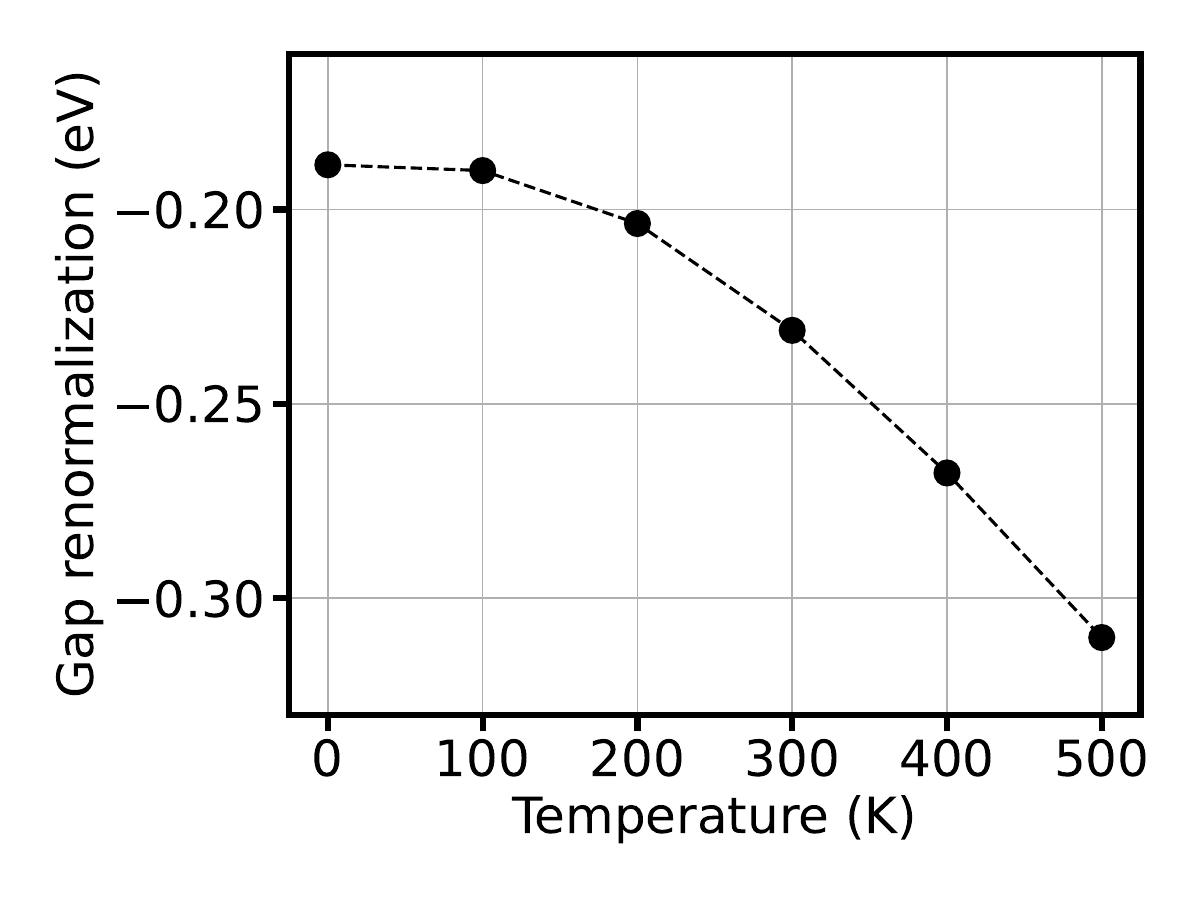}
    \caption{Zero-point and finite-temperature band-gap renormalization due to electron-phonon interactions in $c$-BeB$_2$. The electron-phonon interactions induce a zero-point renormalization of 189 meV and a further temperature-induced reduction of the band gap of 43 meV at 300 K.
    }
    \label{fig:bgr}
\end{figure}

Phonon dispersion calculations reveal no imaginary phonons, indicating that $c$-BeB$_2$ is dynamically stable. The high-frequency dielectric constant, calculated using DFPT, is found to be 13.99. Longitudinal-optical--transverse-optical (LO-TO) splitting at the $\Gamma$-point is observed due to the polar nature of the Be$-$B bonds, resulting in a static dielectric constant considering ionic contribution of 19.28. The computed Born effective charges are +2.2~$e$ on the Be atom, $-$2.1~$e$ on the B$_1$ atom nearest to Be, and $-$0.1~$e$ on the more distant B$_2$ atom.   
The phonon density of states indicates that the majority of the low-frequency optical phonon modes are associated with the neutral boron (B$_2$) while the high-frequency optical modes are associated with the charged boron atom (B$_1$), which is expected due to the stronger interaction. We further characterize the zero-point renormalization and the band gap reduction at finite temperature due to electron-phonon interactions, shown in Fig.~\ref{fig:bgr}. Within the PBE exchange-correlation functional, a zero-point renormalization of 189 meV is seen, which increases as a function of temperature due to stronger phonon population. At 300 K, the renormalization is increased to 232 meV. 

\begin{figure}[!b]
    \centering
    \includegraphics[width=1.0\linewidth]{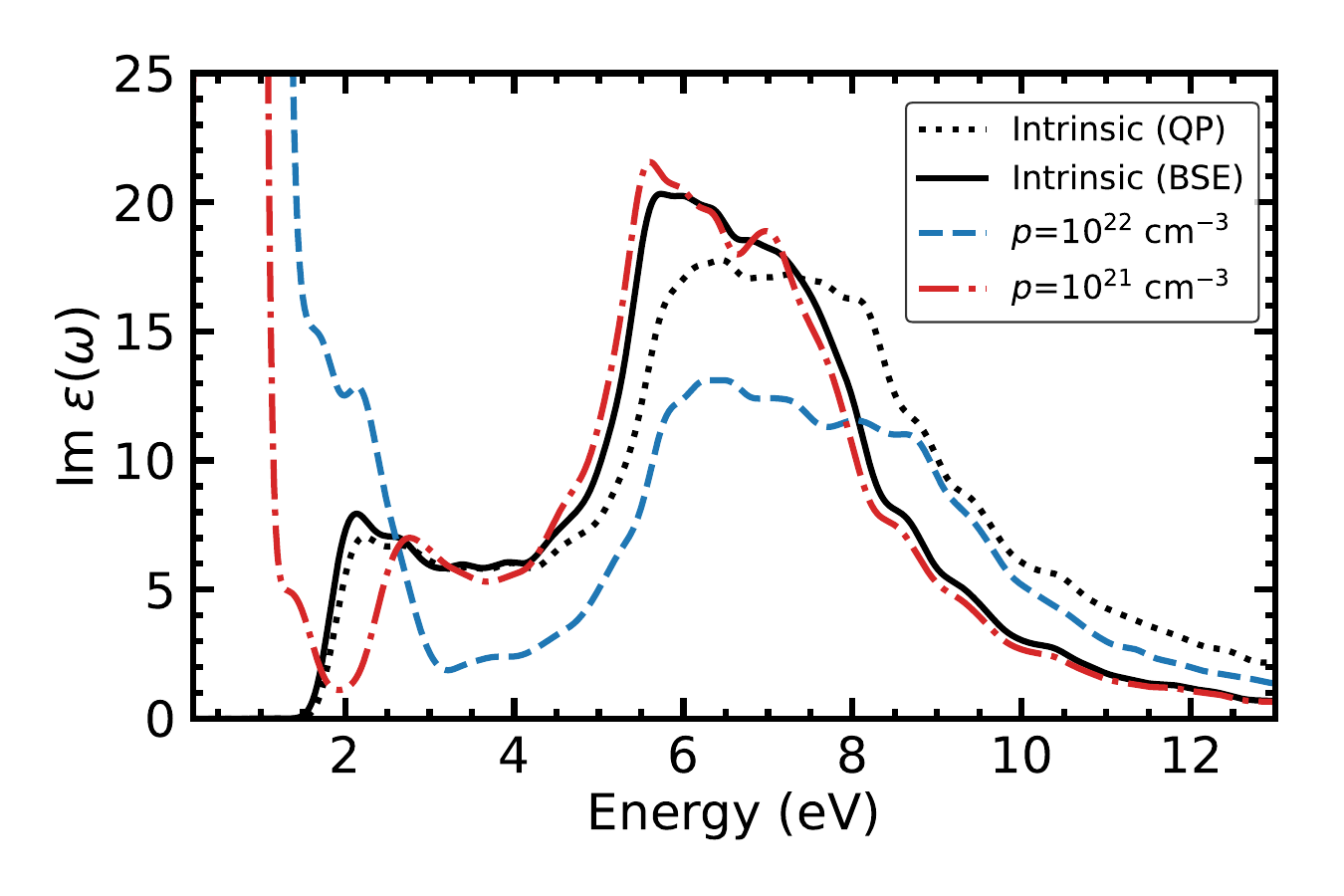}
    \caption{Optical spectra of $c$-BeB$_2$ evaluated with quasiparticle effects only (black dotted line) and with additional inclusion of excitonic effects (black solid line) in the intrinsic material. Carrier concentration-dependent optical spectra including excitonic effects are shown for two doping levels with red dot-dashed and blue dashed lines.  A weak excitonic effect can be seen at the optical absorption onset. Inter-valance band transitions can be clearly seen from the strong optical absorption below a photon energy of 2~eV in the heavily doped material. }
    \label{fig:optics}
\end{figure}

Given the range of the band gaps evaluated using both hybrid functionals and the $GW$ approximation, $c$-BeB$_2$ appears to be a feasible candidate for optoelectronic applications. 
In principle, evaluating optical absorption across an indirect band gap requires accounting for phonon-assisted transitions. However, the difference between the minimum direct gap and the fundamental indirect gap is only 0.28 eV at the $GW$ level. At finite temperature, the broadening of electronic states and spectral features is expected to render direct transitions dominant. Therefore, our analysis focuses on direct optical absorption only.
The calculated optical absorption spectra for both pristine $c$-BeB$_2$ and heavily $p$-type doped samples with carrier concentrations of $p=10^{21}$ cm$^{-3}$ and $p=10^{22}$ cm$^{-3}$ are presented in Fig.~\ref{fig:optics}. 
In the pristine material, a weak excitonic effect is evident by comparing spectra with and without the inclusion of electron-hole Coulomb interactions. The sharp increase in absorption beyond the absorption onset suggests that the material can be an excellent absorber of visible light. In heavily $p$-doped samples, two effects are observed due to the partially empty VBM: First, a gap opening is seen due to the shifting down of the highest occupied level, and second, a strong sub-gap absorption due to inter-valence optical transitions. 
An evaluation of the absorption coefficient shows that in the photon energy range of 2-4 eV, the material exhibits an absorption coefficient exceeding $2\times10^{5}$ cm$^{-1}$ regardless of the impact of the doping level on the band gaps (see Fig.~S4\cite{SM} for the calculated absorption coefficient). 
As a result, despite a strong band gap opening due to Burstein-Moss effect at heavy doping, the material is not a likely candidate for transparent conducting applications. 

\section{Defect formation and electronic transport}

\begin{figure}[!ht]
    \centering
    \includegraphics[width=\linewidth]{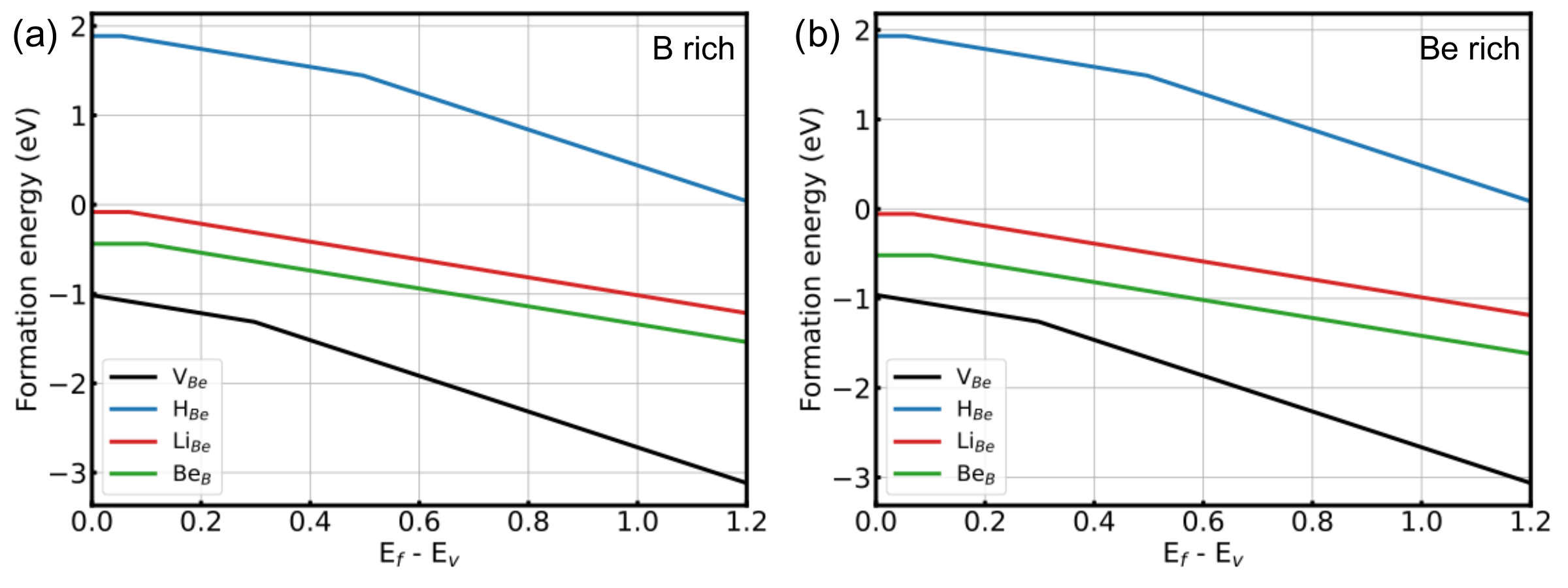}
    \caption{Formation energy of intrinsic (V$_{\text{Be}}$, Be$_{\text{B}}$) and extrinsic (Li$_{\text{Be}}$, H$_\text{Be}$) acceptor-type point defects in $c$-BeB$_2$ as a function of Fermi level (referenced to the valence band maximum $E_V$) under (a) B-rich and (b) Be-rich conditions. As the material is metastable with a low formation enthalpy (-0.054 eV/f.u.), the allowed chemical potential range is narrow, thus defect formation energies are similar under both B-rich and Be-rich conditions. The ionization enegies of the acceptors considered are all less than 100 meV. The Be vacancy, in particular, acts as a shallow acceptor and has a negative formation energy, suggesting potential degenerate $p$-type behavior. 
    }
    \label{fig:defect}
\end{figure}

In order to assess the applicability of $c$-BeB$_2$ as a $p$-type semiconductor, we calculate the defect formation energy for several acceptor point defects as a function of Fermi level and growth conditions, as shown in Fig.~\ref{fig:defect}. 
Due to the low formation energy of $c$-BeB$_2$, the allowed chemical potential range is narrow ($0\geq\Delta\mu_B\geq-0.027$ eV, $0\geq\Delta\mu_{Be}\geq-0.054$ eV). 
As a result, including more stable Be-B compounds will not leave a stable range for the chemical potentials of B and Be for $c$-BeB$_2$. 
In our analysis, we do not consider the formation of other stable Be-B compounds, assuming that the formation of the stable phase is prevented by the growth condition such as epitaxial strain. 
We focus on the demonstration of representative shallow acceptors, intrinsic and extrinsic ones, to show the potential of the material for $p$-type conductivity. 
The intrinsic defects considered are Be vacancy (V$_{Be}$), and Be antisites (Be$_B$). 
Extrinsic point defects considered are H or Li substitutions on the Be site (H$_{Be}$ and Li$_{Be}$). 
Notably, for all acceptor-like point defects considered apart from H$_{Be}$, negative defect formation energies are found. 

While Li$_{Be}$ and Be$_B$ both show an ionization energy of approximately 0.1 eV above the VBM, V$_{Be}$ shows a formation energy of $-1$ eV and stabilizes in the $-1$ charge state at the VBM. 
Therefore, the V$_{Be}$ is expected to be the dominant point defect in $c$-BeB$_2$, acting as a shallow acceptor. 
We note that negative defect formation energies are similarly found in other nitride and oxide compounds, typically suggesting very high doping limits~\cite{deng2021semiconducting} and intrinsic conductivity~\cite{scanlon2011nature}. 
Therefore, we expect $c$-BeB$_2$ to be a potential degenerate $p$-type semiconductor, motivating further investigation of hole transport and potential superconductivity. 

Interestingly, the thermodynamic stability of the BeB$_{2.75}$ composition \cite{VANDERGEEST2014stability,hermann2013binary}, i.e., a Be-deficient polytype compared to $c$-BeB$_2$, suggests a possible relationship between the negative formation energy of Be vacancies with the increased thermodynamic stability of Be-deficient phases. 
Our results provide a qualitative understanding of this connection by attributing it to the strengthening of nearby B-B bonds upon removal of the Be atom (See the analysis and Fig. S5 in the supplemental information\cite{SM}). 
However, the stoichiometry of BeB$_{2.75}$ corresponds to an exceptionally high concentration of Be vacancies ($\sim$ 27\%) in the BeB$_2$ stoichiometry. Moreover, although B adopts a tetrahedral geometry in both polytypes, the two phases have significantly different crystal structures (diamond network in $c$-BeB$_2$ versus icosahedral clusters in BeB$_{2.75}$). Therefore, whether the formation energy of isolated Be vacancies could serve as an indicator for the stability of Be-deficient phases is not immediately obvious and would require further investigations, which are beyond the scope of the present study.

\begin{figure}[!ht]
    \centering
    \includegraphics[width=0.85\linewidth]{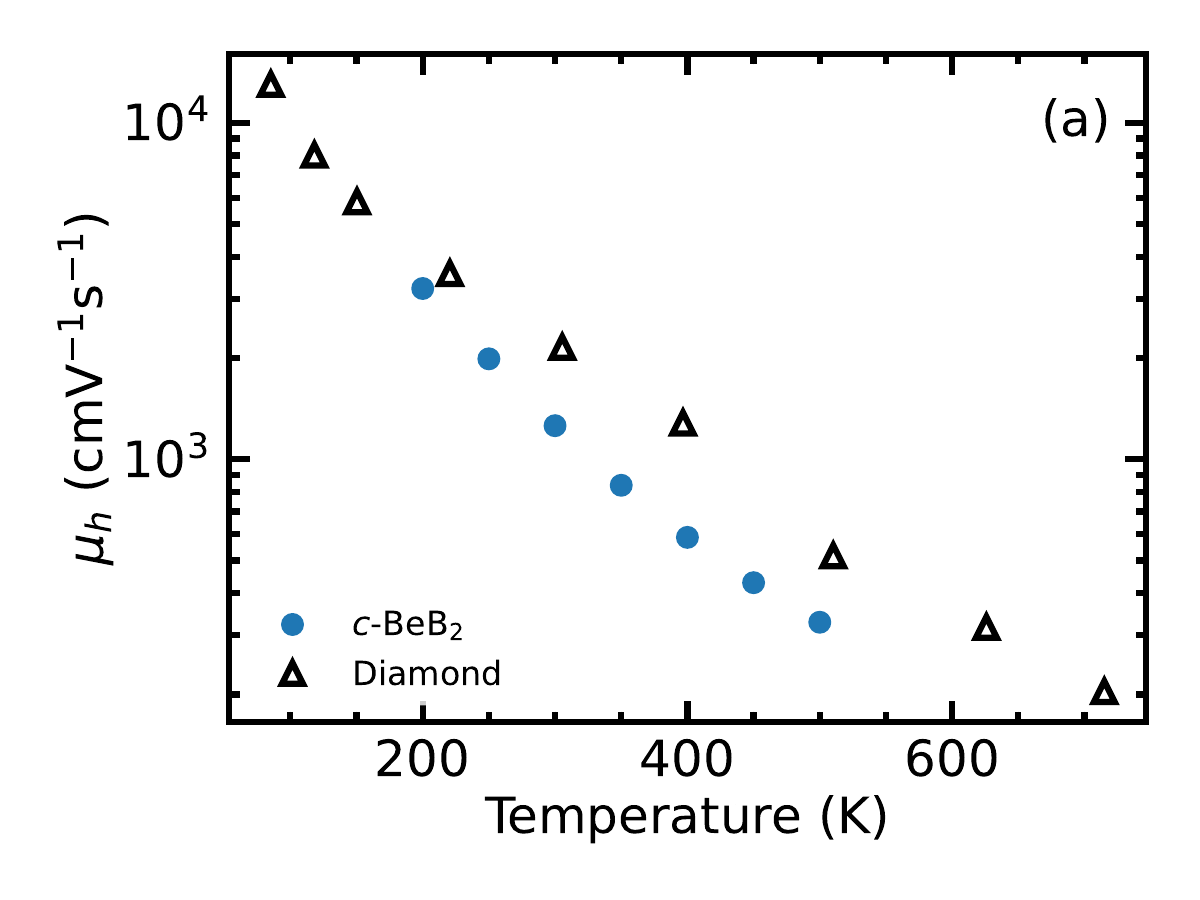}
    \includegraphics[width=0.85\linewidth]{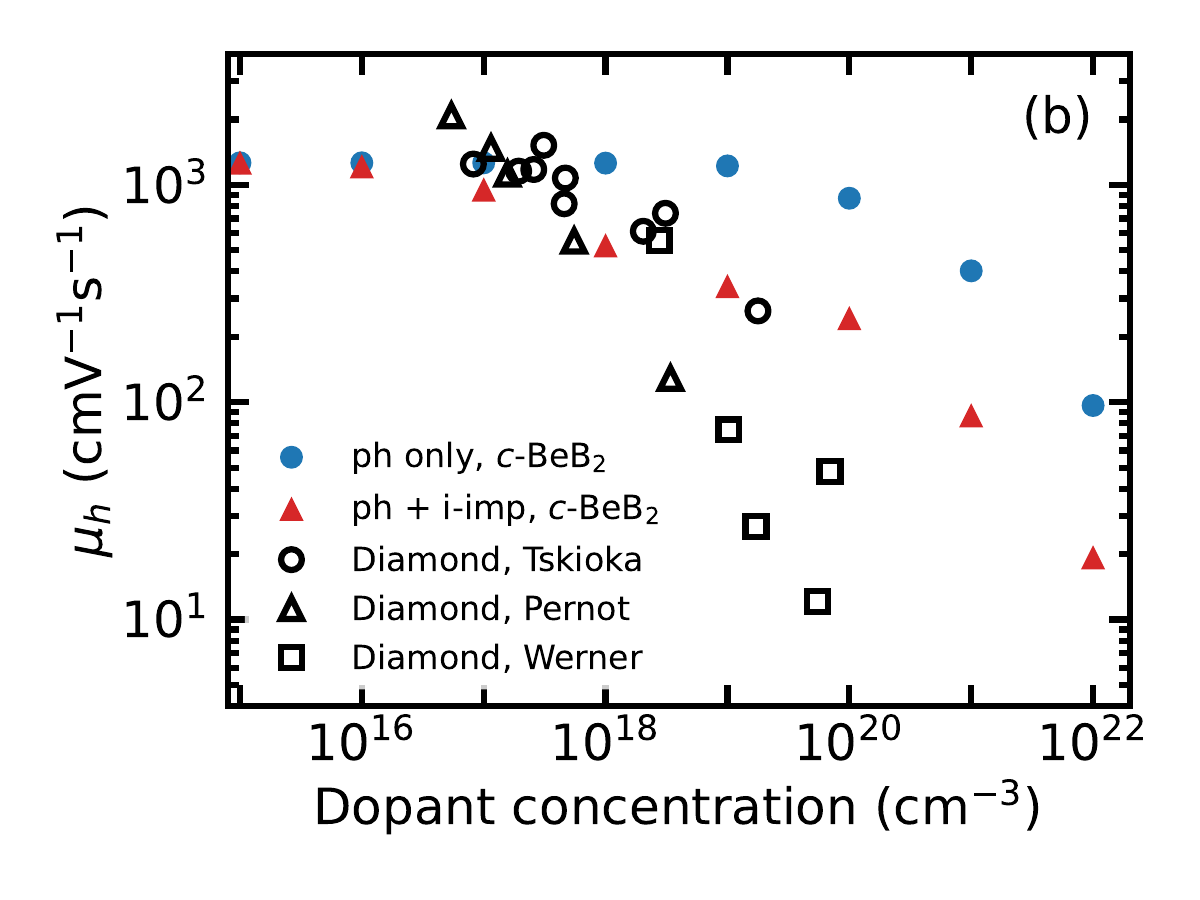}
    \caption{(a) Comparison of calculated temperature dependent mobility in the low doping regime ($p=10^{15}$ cm$^{-3}$) in $c$-BeB$_2$ (blue filled circles) with experimentally measured mobility of diamond from Ref.~\cite{reggiani1983hall} (red filled triangles). $c$-BeB$_2$ exhibits a high hole mobility of 1,259 cm$^2$V$^{-1}$s$^{-1}$ at room temperature (300 K), which is $\sim$58\% of that of diamond.  
    (b) Carrier concentration-dependent hole mobility at 300 K with the inclusion of hole-phonon (ph) scattering (blue filled circles) and both hole-phonon and hole-ionized impurity (i-imp) scattering (red filled triangles). 
    Ionized-impurity scattering strongly limit the hole mobility above a dopant concentration of 10$^{18}$ cm$^{-3}$. 
    Compared to experimental measurements in diamond from Ref.\cite{Tsukioka_2006} (open circles), Ref.~\cite{WERNER1997308} (open triangles) and Ref.~\cite{PhysRevB.81.205203}, a much higher hole mobility is seen in $c$-BeB$_2$ at high dopant concentrations exceeding 10$^{19}$ cm$^{-3}$. }
    \label{fig:mobility}
\end{figure}

Due to the anticipated intrinsic $p$-type conductivity of $c$-BeB$_2$, we next evaluate the hole transport properties in the material with consideration of both hole-phonon scattering and ionized-impurity scattering. 
We show the dependence of the intrinsic hole mobility on temperature and the dependence of hole mobility on dopant concentration at room temperature (300 K) in Fig.~\ref{fig:mobility}. 
Compared to the intrinsic hole mobility of diamond, the intrinsic hole mobility in $c$-BeB$_2$ is lower by approximately 40\%. 
However, the hole mobility still exceeds many known electronic materials such as Si and GaAs. 
As $p$-type conductivity is expected, we examine the hole mobilities from the intrinsic regime up to 10$^{22}$ cm$^{-3}$.
Our results on the carrier-concentration dependence of the hole mobility show that at low doping levels, hole mobility in the material is mainly limited by hole-phonon scattering. 
However, above $p=10^{18}$ cm$^{-3}$, ionized-impurity scattering begins to become the dominant factor, reducing mobility by 60\% at $p=10^{18}$ cm$^{-3}$ and a factor of 5 when reaching a high doping regime such as $10^{21}$ and $10^{22}$ cm$^{-3}$.
Despite the strong impact of ionized-impurity scattering, we find the hole mobilities to remain on the order of 10-100 cm$^2$V$^{-1}$s$^{-1}$ even at very high doping levels, i.e., 86 and 19 cm$^2$V$^{-1}$s$^{-1}$ at $p=10^{21}$ cm$^{-3}$ and $p=10^{22}$ cm$^{-3}$, respectively. 
The resulting electrical conductivities are $1.38\times10^6$ $\Omega^{-1}$m$^{-1}$ and $3.08\times10^6$ $\Omega^{-1}$m$^{-1}$ for the two carrier concentrations.
Therefore, a high electrical conductivity is expected in $c$-BeB$_2$ due to the low acceptor formation and ionization energies and the high hole mobility. 

\section{Superconductivity}

\begin{figure}[!t]
    \centering
    \includegraphics[width=0.48\textwidth]{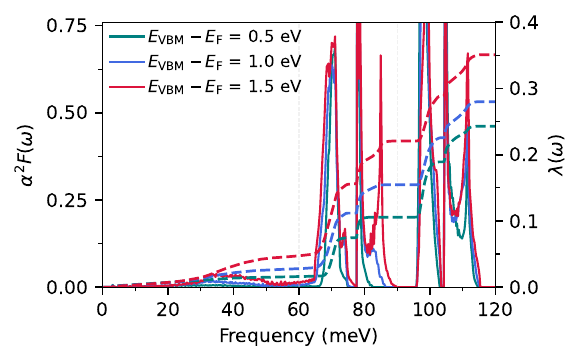}
    \caption{Eliashberg spectral function $\alpha^2F (\omega)$ (solid curves, left axis) and cumulative electron-phonon coupling strength $\lambda (\omega)$ (dashed curves, right axis) for hole-doped $c$-BeB$_2$ with rigid shift of the Fermi level with respect to the valence band maximum of $-$0.5 eV,  $-$1.0 eV, and $-$1.5 eV.}
    \label{fig:a2f}
\end{figure}

\begin{figure*}[!ht]
    \centering
    \includegraphics[width=\textwidth]{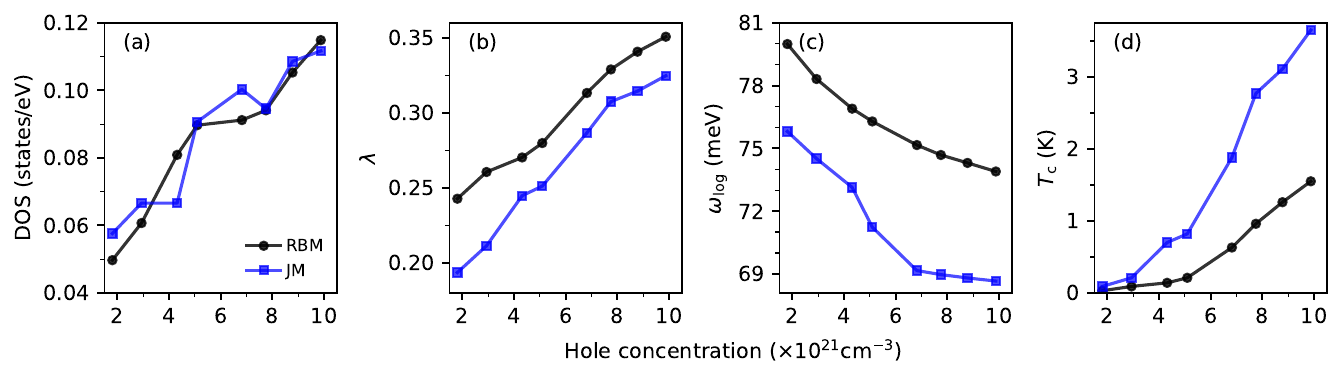}
    \caption{Comparison of the superconducting properties of $c$-BeB$_2$ as a function of hole doping, calculated using a rigid band model (RBM) and a charge-compensated jellium model (JM). Panels show (a) the density of states (DOS) at the Fermi level, (b) the electron-phonon coupling constant ($\lambda$), (c) the logarithmic average phonon frequency ($\omega_{\rm log}$), and (d) the superconducting critical temperature ($T_{\rm c}$) estimated using the McMillan formula with a semi-empirical Coulomb parameter $\mu^*$ = 0.1.} 
    \label{fig:jellium}
\end{figure*}

To examine the superconducting properties of hole-doped $c$-BeB$_2$, we first consider a rigid-band model (RBM), where the Fermi level is shifted in the range of $-$0.5~eV to $-$1.5~eV, corresponding to carrier concentrations of $1.8 \times 10^{21}$ cm$^{-3}$ to $9.8 \times 10^{21}$ cm$^{-3}$. The selected doping levels are comparable to those considered in studies of superconductivity in B-doped diamond~\cite{ekimov2004superconductivity,lee2004superconductivity,blase2004role,Giustino2007Electron,Costa2021}, Si~\cite{bustarret2006superconductivity,bourgeois2007superconductivity} and SiC~\cite{Ren2007,Kriener2008superconductivity,Margine2008,noffsinger2009origin}. 
The calculated Eliashberg spectral function, $\alpha^2F(\omega)$, and the cumulative electron-phonon coupling strength, $\lambda(\omega)$, for three hole-doping levels are shown in Fig.~\ref{fig:a2f}. The spectral function can be divided into three regions: below 60 meV, between 60$-$90~meV, and above 90~meV. Across all doping levels, the modes below 60~meV, associated mainly with acoustic phonons, have a minimal effect on the total electron-phonon coupling strength, accounting for only about 5 - 10\%. This result is qualitatively similar to the case of B-doped Si~\cite{bourgeois2007superconductivity} and B-doped diamond~\cite{blase2004role,Giustino2007Electron,Costa2021}, where acoustic phonons have negligible coupling, but unlike B-doped SiC, where these modes dominate ($\sim$ 70\%) ~\cite{noffsinger2009origin,Margine2008}. The mid-frequency optical modes in the 60$-$90~meV region play an increasingly important role with increasing doping. Their contribution to the total $\lambda$ rises from approximately 37\% to 50\% at $-$0.5 eV and $-$1.5~eV, respectively. In contrast, the high-frequency modes above 90~meV exhibit an opposite trend, their contribution dropping from 56\% to 37\% over the same doping range. Consequently, the progressive increase in $\lambda$ from 0.24 to 0.35 with doping is primarily driven by the enhanced coupling in the 60$-$90~meV range, which directly influences the superconducting properties. 

To further investigate the effect of doping on the phonon dispersion, we also adopt a charge-compensated jellium model (JM). Fig.~S6~\cite{SM} compares the phonon dispersion of the undoped system with that of a doped system within the JM for a doping level of $9.8 \times 10^{21}$ cm$^{-3}$ corresponding to a Fermi level shift of 1.5~eV below the VBM. The JM leads to noticeable changes in the optical phonon branches, particularly at higher frequencies. Notably, the LO-TO splitting vanishes, which can be attributed to the metallic screening arising from partially occupied electronic states. This screening suppresses the long-range dipole-dipole interactions responsible for LO-TO splitting, leading to a softening of the optical phonons, especially near the $\Gamma$ point. Such phonon softening significantly affects the electron-phonon coupling.
 
In Fig.~\ref{fig:jellium}, we compare the superconducting properties of doped $c$-BeB$_2$ as a function of hole concentration, calculated using both the RBM and JM. Both approaches capture the same qualitative trend, an increase in the DOS at the Fermi level with increasing doping concentration. In conventional BCS-type superconductors, an enhanced DOS at the Fermi level strengthens the electron-phonon coupling constant $\lambda$, which in turn leads to an increase in the superconducting $T_{\rm c}$. Here, the $T_{\rm c}$ is estimated using the McMillan formula~\cite{McMillan1968} with a Coulomb pseudopotential $\mu^*$ = 0.1. Although the JM yields slightly smaller $\lambda$ values, the predicted $T_{\rm c}$ is more than double that of the RBM at high doping levels. This enhancement is primarily attributed to the softening of optical phonons within the JM, which reduces the logarithmic average phonon frequency ($\omega_{\rm log}$), thereby amplifying the $T_{\rm c}$ despite modest differences in $\lambda$.

Finally, we compare the superconducting properties of hole-doped $c$-BeB$_2$ with those of other heavily hole-doped covalent semiconductors. At the highest doping level, the RBM yields $\lambda=0.35$ and $\omega_{\rm log}=73.9$~meV, resulting in a predicted $T_{\rm c}=1.55$~K for $\mu^*=0.1$. In comparison, the JM gives a slightly lower $\lambda=0.32$ and $\omega_{\rm log}=68.7$~meV, leading to a higher $T_{\rm c}=3.65$~K. These values are comparable to those found in other heavily hole-doped covalent systems. For instance, in B-doped diamond $\lambda$ is $\sim 0.39$~\cite{Xiang2004}, with experimentally observed $T_{\rm c}$ values ranging from 2.3~K to 4~K ~\cite{ekimov2004superconductivity}. B-doped Si shows much weaker coupling, with $\lambda$ between 0.1 and 0.2 and $T_{\rm c}$ around 0.35~K~\cite{bustarret2006superconductivity}. B-doped SiC falls between these cases, with $\lambda$ values of 0.3$-$0.4 and $T_{\rm c}$ around 1.5~K~\cite{Ren2007,Margine2008,noffsinger2009origin}. These comparisons suggest that hole-doped $c$-BeB$_2$ exhibits superconducting properties on par with other heavily doped covalent semiconductors.

\section{Conclusion}

In this work, we provide a comprehensive first-principles investigation of $c$-BeB$_2$, a metastable zinc blende boride formed due to electron donation of Be to B. Although the material is metastable compared to other BeB$_2$ polytypes, its close lattice match with established cubic and hexagonal materials suggests a potential route for epitaxial stabilization. The calculated phonon dispersion confirms the dynamical stability of the metastable cubic phase under ambient conditions. 

We evaluate the electronic and optical properties of $c$-BeB$_2$. 
The material is found to be an indirect semiconductor with a fundamental gap on the order of 1.5~eV with very low effective masses for hole transport. 
Defect calculations reveal that Be vacancy, which is a shallow acceptor defect with negative formation energy, dominates defect formation in the material, therefore making it intrinsically degenerate $p$-type. We further predict a high intrinsic hole mobility of 1,259 cm$^2$s$^{-1}$V$^{-1}$ at room temperature, considering both hole-phonon and hole-ionized-impurity scattering. The mobility of $p$-type $c$-BeB$_2$ is found to be mainly limited by hole-ionized-impurity scattering and remains high even for very high doping levels. Furthermore, we find that heavily doped $p$-type $c$-BeB$_2$ is a low-temperature superconductor, exhibiting a $T_{\rm c}$ of up to 3.55 K based on Eliashberg theory, comparable to other heavily doped covalent semiconductors such as diamond, Si, and SiC.

Our findings establish $c$-BeB$_2$ as a $p$-type boride semiconductor with potential electronic, optoelectronic, and superconducting applications. While bulk synthesis is expected to be challenging due to its metastability, our work motivates further efforts to realize this metastable phase in thin-film form and explore its functional properties. 

\section*{acknowledgment}

The work is supported as part of the Computational Materials Sciences Program funded by the U.S. Department of Energy, Office of Science, Basic Energy Sciences, under Award No. DE-SC0020129. Computational resources were provided by the National Energy Research Scientific Computing Center, which is supported by the Office of Science of the U.S. Department of Energy under Contract No. DE-AC02-05CH11231, and the Texas Advanced Computing Center (TACC) at The University of Texas at Austin (http://www.tacc.utexas.edu) through the ACCESS program under Award No. TG-DMR180071.  

\bibliography{refs}

\end{document}